\documentclass[%
 aip,
 jcp,
 amsmath,amssymb,
 reprint
]{revtex4-1}
\usepackage[english]{babel}
\usepackage[percent]{overpic}
\usepackage{orcidlink}
\usepackage{array} 
\usepackage{amsmath}
\usepackage{mathbbol}
\usepackage{amssymb}
\usepackage{mathrsfs}
\usepackage{siunitx}
\usepackage{dutchcal}
\usepackage{tikz}
\usepackage{makecell}
\usepackage[acronym,nomain,hyperfirst=false]{glossaries}
\usepackage{braket}
\usepackage{xspace}% used for space after commands
\usepackage{amsmath}% used for position of tilde 
\usepackage{multirow}
\usepackage{booktabs}
\usepackage{csquotes}
\usepackage{xcolor}
\hypersetup{colorlinks=true, allcolors=blue}
\usepackage[capitalise]{cleveref}

\usetikzlibrary{positioning,calc,arrows.meta}
\usetikzlibrary{shapes.geometric, arrows}
\usetikzlibrary{arrows.meta}

\definecolor{steelblue}{HTML}{4682B4}
\definecolor{firebrick}{HTML}{B22222}
\definecolor{excitinggold}{HTML}{F4CB3B}
\definecolor{gsblue}{HTML}{7FA6D6}
\definecolor{processcyan}{HTML}{6FC2B0}

\newcommand{\ie}{{\it i.e.}, }
\newcommand{\exciting}{\texttt{exciting}\xspace} % exciting 
\makeatletter
\def\@email#1#2{%
 \endgroup
 \patchcmd{\titleblock@produce}
  {\frontmatter@RRAPformat}
  {\frontmatter@RRAPformat{\produce@RRAP{*#1\href{mailto:#2}{#2}}}\frontmatter@RRAPformat}
  {}{}
}%

\begin{document}
\title{First-principles description of pumped inelastic X-ray scattering: example of K-edge RIXS in graphite}

\author{Elias Richter\orcidlink{0009-0008-2773-1248}}
\email{elias.richter@physik.hu-berlin.de}
\affiliation{Department of Physics and CSMB, Humboldt-Universität zu Berlin, Berlin, Germany}
\author{Benedikt Maurer\orcidlink{0000-0001-9152-7390}}
\affiliation{Department of Physics and CSMB, Humboldt-Universität zu Berlin, Berlin, Germany}
\author{Claudia Draxl\orcidlink{0000-0003-3523-6657}}
\affiliation{Department of Physics and CSMB, Humboldt-Universität zu Berlin, Berlin, Germany}
\affiliation{European Theoretical Spectroscopy Facility (ETSF)}

\begin{abstract}
We present an \textit{ab initio} framework for predicting resonant inelastic X-ray scattering (RIXS) in optically pumped materials. Our methodology is based on the Kramers-Heisenberg formula for the double-differential cross section formulated using the results of the Bethe-Salpeter equation (BSE) from many-body perturbation theory. To extend this approach to the time domain, we incorporate non-equilibrium charge-carrier distributions obtained from real-time, time-dependent density-functional theory (RT-TDDFT). Generalizing the RIXS implementation with respect to arbitrary polarizations, allows us to consider different orientations of incoming and outgoing light. We demonstrate our method's capabilities by studying RIXS at the K-edge of graphite for various non-equilibrium charge-carrier distributions, representing different delay times after optical pumping. Our results reveal angular dependencies in $\pi$- and $\sigma$-orbital-derived spectral regions, in good agreement with experiment. 
\end{abstract}

\maketitle

\section{Introduction}

Resonant inelastic X-ray scattering (RIXS) is a state-of-the-art spectroscopy technique that provides access to electronic excitations over a wide spectral range \cite{Ament2011}. Owing to its bulk-, element-, and orbital-sensitivity, RIXS has emerged as a powerful tool to study excitations in solids \cite{Ament2011,Wang2017,Ghiringhelli2005}. In particular, RIXS enables access to momentum- and polarization-resolved spectral features, offering insights into the underlying electronic structure as well as energy, symmetry, and orbital character of electronic states. This simultaneous access to multiple properties distinguishes RIXS from other methods, which typically probe only a subset of these properties. In this context, theory can provide insight into the details of the RIXS process, yielding a detailed understanding of how electronic excitations and their spectral signatures emerge from many-body interactions. 

While early experimental work \cite{Ma1992} triggered interest in theoretical work, first-principles approaches to RIXS were long limited to the special case of non-resonant x-ray emission spectroscopy (XES) within the independent particle approximation (IPA) \cite{Johnson1994,Jia1996,Strocov2004,Strocov2005}. Extensions incorporating explicit core-hole effects, together with an on-site Coulomb correction to enhance the description of localized states~\cite{Magnuson2010}, led to partial improvement. For molecules, a variety of electronic-structure methods have been used to describe RIXS in small systems. These include wavefunction-based approaches such as multi-configurational self-consistent field~\cite{Josefsson2012} and configuration interaction theories~\cite{Maganas2017}, as well as equation-of-motion coupled-cluster and damped-response formulations ~\cite{Faber2019,Faber2020,Nanda2020,Nanda2020-2,Schnack-Petersen2023}. In addition, Green’s-function-based approaches like algebraic diagrammatic construction~\cite{Rehn2017} and density-based methods within linear-response time-dependent density-function theory (LR-TDDFT)  ~\cite{Nascimento2022,Roychoudhurry2022,Vitols2025} have been adopted. However, excitonic effects, which are important for both core and valence excitations, in particular for semiconducting systems, were only considered more recently in RIXS implementations based on the Bethe-Salpeter equation (BSE) of many-body perturbation theory (MBPT) \cite{Shirley1998,Shirley2000,Shirley2001,Vinson2012,Vinson2017,Vinson2019,Geondzhian2018,Vorwerk2020,Vorwerk2022}. While most of them relied on the pseudopotential approximation, the approach by Vorwerk and coworkers \cite{Vorwerk2020,Vorwerk2022} is based on the full-potential code \exciting \cite{Exciting2014,Exciting2026}. 

Time-resolved RIXS (tr-RIXS) extends the technique to access the dynamics of excited states following an optical pump in ultrafast pump-probe spectroscopy \cite{Gel'mukhanov1994,Carlisle1995,Beye2013,Dell'Angela2013,Ulstrup2015}. Experimental studies remain limited but have demonstrated the capability to probe transient changes in correlated materials \cite{Dean2016,Mazzone2021,Paris2021,Mitrano2024,Thielemann2024,Xu2025}, where it can track pump-induced changes in charge, spin, orbital, and lattice excitations. More recent works highlight the potential of tr-RIXS for investigating non-equilibrium phenomena \cite{Monney2023,Mitrano2024,Xu2025}. Theoretical studies have also indicated the importance of electron-lattice dynamics onto tr-RIXS~\cite{Monney2023,Freibert2024,Dashwood2021}. Overall, the theoretical modeling of tr-RIXS requires not only an accurate description of electron-hole correlations, but also a framework capable of capturing non-equilibrium effects induced by the pump.

In this work, we present an \textit{ab initio} framework for polarization- and time-resolved RIXS calculations, combining RT-TDDFT and BSE. It represents an extension of the workflow proposed in Ref. \cite{Vorwerk2020,Vorwerk2021,Vorwerk2022}. We demonstrate our approach for graphite, for which experimental studies are available~\cite{Malvestuto2022,Malvestuto2025,Malvestuto2026}. Due to its prototypical layered structure and strong electronic anisotropy, graphite exhibits complex excitation features and a clear separation between out-of-plane $\pi$ and in-plane $\sigma$ states, leading to strongly polarization-dependent excitation spectra. The combination of strong in-plane covalent bonding and weak inter-layer van der Waals interactions results in semi-metallic electronic properties and make it an ideal system for studying fundamental excitations in solids \cite{Dresselhaus2002}, and in particular their directional dependence. Excitonic effects need to be accounted for in both valence \cite{Magnuson2010,Trevisanutto2010} and core-level excitations \cite{Shirley1998,Unzog2022}. Earlier theoretical and experimental studies have also shown that RIXS at the carbon K-edge of graphite is highly sensitive to symmetry selection rules and polarization, allowing one to resolve characteristic $\pi$- and $\sigma$-state excitations already at the level of the electronic structure \cite{Carlisle1995,VanVeenendaal1997,Carlisle1999,Shirley2000,Shirley2001}. Similar results have been reported for graphene \cite{Zhang2012}. Moreover, owning the pronounced anisotropy of the material, the scattering geometry and participating polarizations have a decisive impact and must be consistently accounted for~\cite{Yasui2006,XuanGao2013}.

The structure of this paper is as follows. \cref{Sec:Theory} presents the theoretical framework underlying polarization- and time-resolved RIXS and its relationship to BSE. \cref{Sec:Implementation} describes the implementation of our approach within an \textit{ab initio} workflow. \cref{Sec:ComputationalDetails} provides computational details used in the calculations. Finally, in \cref{Sec:Results}, we first discuss the calculated absorption spectra of graphite, before analyzing equilibrium RIXS features and exploring the non-equilibrium response following optical excitation.

\section{Theory} \label{Sec:Theory}

\begin{figure}[h]
    \centering
    \begin{overpic}[width=0.45\textwidth]{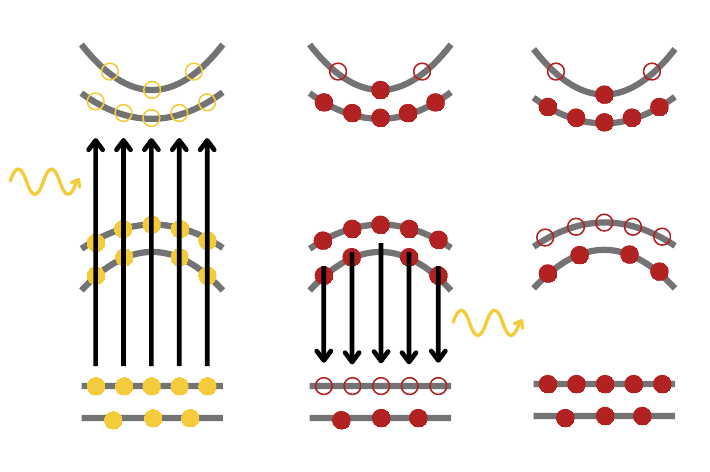}
        \put(4,44.5){$\omega_1$}  
        \put(65.5,24){$\omega_2$}
        \put(6,20){$t^{(1)}_{\lambda_c}$}
        \put(35,20){$t^{(2)}_{\lambda_0,\lambda_c}$}
    \end{overpic}
    \caption{Schema of the RIXS process. Starting from the ground state, the absorption of an incoming X-ray photon with energy $\omega_1$ (left panel) yields an intermediate many-body state (middle panel). Subsequent de-excitation through emission of an X-ray photon $\omega_2$ leads to the final state containing an electron-hole pair with the hole in the valence region (right panel). Filled and open circles represent occupied and unoccupied states, respectively. Black lines represent transitions and yellow wiggly lines represent X-ray absorption and emission.}
    \label{fig:rixs-schema}
\end{figure}

Resonant inelastic X-ray scattering (RIXS) is a photon-in–photon-out process that proceeds via two sequential steps. The microscopic mechanism is illustrated schematically in Fig.~\ref{fig:rixs-schema} and can be described as follows \cite{Kotani2001,Ament2011}: An incident X-ray photon with energy $\omega_1$ is absorbed by the sample, initially in its ground-state $|0\rangle$. This absorption excites a core electron into the conduction band, leaving a core hole behind and creating an intermediate state $|n\rangle$. The core hole can then be filled by a valence electron yielding the emission of a photon $\omega_2$ and leaving a hole in the valence band. Thus, the final excited state $|f\rangle$ consists of this hole and an excited electron in the conduction band. This final state corresponds to that of an optical excitation, reflected by the small energy loss. Although both the absorbed and emitted photons lie in the X-ray regime, their energy difference $\omega = \omega_1-\omega_2$ typically amounts to only a few eV. Following the generalized Kramers-Heisenberg formula \cite{KramersHeisenberg} limiting onto one electron-hole pair in the intermediate state and neglecting non-resonant inelastic X-ray scattering (NRIXS), the RIXS process of an ingoing photon with energy $\omega_1$ and polarization $\mathbf{e}_1$ into an outgoing photon with energy $\omega_2$ and polarization $\mathbf{e}_2$ is described in terms of the double-differential cross section (DDCS), given by
\begin{equation}
\begin{aligned}
    \frac{d^2\sigma}{d\Omega_2 d\omega_2}=\alpha^4 \left( \frac{\omega_2}{\omega_1} \right) \sum_f & \left| \sum_n\frac{ \langle f|\hat{T}^\dagger(\mathbf{e}_2)|n\rangle \langle n|\hat{T}(\mathbf{e}_1)|0\rangle}{\omega_1-E_n}  \right|^2 \\
    &\times \delta \left(E_f+\omega_2 -\omega_1\right).
\label{eq:RixsDdcs}
\end{aligned}
\end{equation}
Here, $E_0=0$ and $E_f$ describe the initial and final state, respectively, and the transition operator is approximated by $\hat{T}(\mathbf{e})\approx\sum_j \mathbf{e}\cdot \mathbf{p}_j$, where $\mathbf{p}_j$ is the momentum operator of the j-th electron. Since the intermediate state contains a core hole, $\mu \mathbf{k}$, and an excited electron in a conduction state, $c \mathbf{k}$, it can be expressed as $|n\rangle =|c\mu \mathbf{k}\rangle$ and the final state as $|f\rangle =|c'v \mathbf{k}'\rangle$ with energies $E_n$ and $E_f$, respectively.

To describe the excited states in our framework, we employ the BSE of MBPT \cite{Salpeter1951,Hedin1965}, which is the state-of-the-art spectroscopy technique beyond the independent particle picture. Thus, it explicitly captures excitonic effects, which are essential for an accurate description of both valence and core-level excitations \cite{Strinati1988,Rohlfing1998,Rohlfing2000,Onida2002}. Within the Tamm–Dancoff approximation, it can be cast into an effective eigenvalue problem,
\begin{equation}
    \sum_{o'u'\mathbf{k}'}H^\mathrm{BSE}_{ou\mathbf{k},o'u'\mathbf{k}'}A^\lambda_{o'u'\mathbf{k}'}=E^\lambda A^\lambda_{o'u'\mathbf{k}'}\,,
\end{equation}
where $A^\lambda_{ou\mathbf{k}}$ is the excitonic eigenvector, $E^\lambda$ is the corresponding exciton energy, and $\mathbf{k}$ indicates points in the reciprocal space. Here, $o$ and $u$ denote occupied and unoccupied single-particle states, respectively. Depending on the excitation under consideration, $o$ corresponds either to a valence state $v$ or a core state $\mu$, while $u$ labels a conduction state $c$.

Describing the excited states in terms of the BSE, allows for an interpretation of the RIXS process in terms of pathways~\cite{Vorwerk2020,Vorwerk2022}, as already presented in \cref{fig:rixs-schema}. The initial X-ray excitation process from the ground state $|0\rangle$ to an intermediate state $|\lambda_c\rangle$ is described by the core oscillator strength 
\begin{equation}
t^{(1)}_{\lambda_c} =\sum_{c\mu\mathbf{k}}A^{\lambda_c}_{c\mu\mathbf{k}} \left[\mathbf{e}_1 \cdot \mathbf{P}_{c\mu\mathbf{k}}\right]\:,
\label{eq:t1}
\end{equation}
the subsequent de-excitation process from $|\lambda_c\rangle$ onto to the final state $|\lambda_0\rangle$ by the excitation pathway
\begin{equation}
    t^{(2)}_{\lambda_0,\lambda_c} =\sum_{cv\mathbf{k}} \sum_\mu A^{\lambda_0}_{cv\mathbf{k}} \left[\mathbf{e}_2^* \cdot \mathbf{P}_{\mu v\mathbf{k}}\right] \left[A^{\lambda_c}_{c\mu\mathbf{k}}\right]^*\: .
\label{eq:t2}
\end{equation}
Combining the two, allows to introduce the RIXS oscillator strength $t^{(3)}_{\lambda_0}(\omega_1)$ 
\begin{equation}
    t^{(3)}_{\lambda_0}(\omega_1) =
     \sum_{\lambda_c}\frac{ t^{(2)}_{\lambda_0,\lambda_c} \: t^{(1)}_{\lambda_c}}{\omega_1-E^{\lambda_c}+i\eta} \:,
\label{eq:t3}
\end{equation}
and to write the DDCS as
\begin{equation}
    \frac{d^2\sigma}{d\Omega_2 d\omega_2} =
    \alpha^4 \left( \frac{\omega_2}{\omega_1} \right)\mathrm{Im} \sum_{\lambda_0} \frac{\left| t^{(3)}_{\lambda_0}(\omega_1) \right|^2}{(\omega_1-\omega_2)-E^{\lambda_0}+i\eta} \:.
\label{eq:RixsDdcsCompact}
\end{equation}
To evaluate this equation, one can employ results from two BSE calculations. A first calculation is performed to obtain the core excitation at a specific edge, enabling the calculation of the core absorption oscillator strength $t^{(1)}$. The second calculation describes the final state corresponding to an optical excited state. Combining both allows for constructing the excitation pathway $t^{(2)}$. The RIXS oscillator strength $t^{(3)}(\omega_1)$ can then be calculated for an explicit choice of excitation energy $\omega_1$, enabling the final construction of the RIXS DDCS.

In case of tr-RIXS, the process must be extended to include the dynamical effects induced by an additional optical pump laser. For this purpose, the RT-TDDFT framework is employed to propagate the Kohn-Sham system in time. The so obtained non-equilibrium occupation factors of valence and conduction bands, $f_{o\mathbf{k}}(t)$ and $f_{u\mathbf{k}}(t)$, respectively, serve as input for subsequent BSE calculations, as will be described in more detail in \cref{Sec:Implementation}.

\section{Implementation} \label{Sec:Implementation}

To recall, RIXS calculations as implemented in the package \textsf{BRIXS}  ~\cite{Vorwerk2020,Vorwerk2021,Vorwerk2022,git-brixs,git-pybrixs} following \cref{eq:RixsDdcsCompact} require explicit access to BSE eigenvectors, $A^{\lambda_c}_{c\mu\mathbf{k}}$ and $A^{\lambda_0}_{cv\mathbf{k}}$, and eigenvalues $E^{\lambda_c}$ and $E^{\lambda_0}$ for core-conduction and valence-conduction transitions, respectively, as well as the corresponding momentum matrix elements $P_{c\mu\mathbf{k}}$ and $P_{cv\mathbf{k}}$, which enter the pathway matrices $t^{(1)}$, $t^{(2)}$, and $t^{(3)}(\omega_1)$ for explicit choices of excitation energy $\omega_1$ and energy loss $\omega=\omega_1-\omega_2$. Then, \textsf{pyBRIXS} calculates the RIXS cross section and corresponding spectra. 

The RIXS DDCS (\cref{eq:RixsDdcs}) depends on both the polarization $\mathbf{e}_1$ of the incoming and $\mathbf{e}_2$ of the outgoing X-ray photons. In the formalism, this shows up, respectively, in the core absorption oscillator strength $t^{(1)}$ (\cref{eq:t1}) and the excitation pathway $t^{(2)}$ (\cref{eq:t2}). To account for the most general case, we have extended the existing implementation, which so far only allowed for equal choice, $\mathbf{e}_1=\mathbf{e}_2$. 

Furthermore, we extended the method to obtain time-resolved RIXS (tr-RIXS) spectra. For this case, \exciting's RT-TDDFT framework \cite{RTTDDFT} is employed to simulate the propagation of the system for a given pump setup and obtain the corresponding electron and hole distributions \cite{Rossi2025,qiao2026}. The time-dependent occupations $f_{n\mathbf{k}}(t)$ define the occupation factors
\begin{equation}
    F_{ou\mathbf{k}}(t)
    =
    \sqrt{f_{o\mathbf{k}}(t)-f_{u\mathbf{k}}(t)}.
\end{equation}
which allows us to introduce time dependence into the RIXS simulations in two ways. First, the obtained non-equilibrium occupation numbers are used in subsequent BSE calculations for valence and core-level excitations. They modify the screened Coulomb interaction and, through $F_{ou\mathbf{k}}(t)$, also the electron–hole interaction kernel, which is symmetrized accordingly and thus remains Hermitian. Consequently, both enter the BSE eigenvalues and eigenvectors. Second, the same factors enter the RIXS pathways explicitly: the eigenvector components in \cref{eq:t1,eq:t2} are replaced by $A^{\lambda_c}_{c\mu\mathbf{k}}\rightarrow
F_{\mu c\mathbf{k}}(t)A^{\lambda_c}_{c\mu\mathbf{k}}$
and $A^{\lambda_0}_{cv\mathbf{k}}\rightarrow
F_{vc\mathbf{k}}(t)A^{\lambda_0}_{cv\mathbf{k}}$. This accounts for Pauli blocking. Since $t^{(3)}$ is a coherent combination of $t^{(1)}$ and $t^{(2)}$, the occupation factors modify both the weights and the interference of the individual RIXS pathways.

In \cref{fig:workflow}, we display the full workflow, covering the unpumped as well as the pumped cases. The dielectric tensors for both valence and core-level excitations are computed by the BSE, which contain the full information on all polarization directions. From the core excitation, we also obtain the core oscillator strength $t^{(1)}$, related to the polarization of the incoming light beam $\mathbf{e}_1$, which can be chosen to reflect a specific setup. The excitation pathway $t^{(2)}$ combines input from both the core and valence BSE results after choosing the polarization $\mathbf{e}_2$ of the outgoing beam. The RIXS oscillator strength $t^{(3)}_{\lambda_0}(\omega_1)$ is then computed for a range of excitation energies $\omega_1$, reflecting the selected core edge. Finally, the DDCS and the RIXS spectra are obtained.

\begin{figure}[h]
    \centering
    \begin{tikzpicture}[
  scale=0.45, transform shape,
  every node/.append style={font=\sffamily\huge, align=center,
    execute at begin node={\setlength{\baselineskip}{15pt}}},
  arrow/.style={thick,->,>=Stealth},
  link/.style={thick},
  startstop/.style={rectangle, rounded corners=4pt, minimum height=12mm, text width=6cm, draw=black, fill=gsblue!60},
  process/.style={rectangle, minimum height=18mm, text width=8.0cm, draw=black, fill=processcyan!30},
  spectra/.style={rectangle, rounded corners=6pt, minimum height=12mm, text width=7cm, draw=black, fill=excitinggold!70}
]

% ---------- Nodes ----------
\node (gs) [startstop] {DFT ground state};

\node (rttddft) [process, below=15mm of gs] {RT-TDDFT \\ optional: pumped case};

\node (bse-core)    [spectra,    below left= 15mm and -10mm of rttddft] {BSE: core spectra};
\node (bse-val)    [spectra,    below right=15mm and -10mm of rttddft] {BSE: optical spectra};

\node (pathway) [process, below =37mm of rttddft] {RIXS pathways \\ $t^{(1)}$, $t^{(2)}$, $t^{(3)}(\omega_1)$};
\node (ddcs) [spectra, below=18mm of pathway]    {RIXS cross section};

\draw[arrow,dashed] (gs) -- (rttddft);
\draw[arrow,dashed] (rttddft) -- (bse-val);
\draw[arrow,dashed] (rttddft) -- (bse-core);
\draw[arrow,dashed] (rttddft) -- (pathway);
\draw[arrow] (gs) -| (bse-val);
\draw[arrow] (gs) -| (bse-core);

\draw[arrow] (bse-core) |- (pathway);
\draw[arrow] (bse-val) |- (pathway);

\draw[arrow] (pathway.south)              -- node[right, pos=0.5] {choice of $\mathbf{e}_1$, $\mathbf{e}_2$} (ddcs.north) ;

\end{tikzpicture}
    \caption{Workflow representing RIXS and tr-RIXS calculations: A DFT calculation serves as starting point\footnote{Ideally, one would perform a $GW$ calculation on top of DFT to obtain the quasi-particle energies. We don't include this step here for two reasons. Core states would require a self-consistent $GW$ scheme, which is not not available in \exciting yet. Second, also RT-TDDFT calculations cannot be carried out on top of $GW$ required for the pumped case.} for both optical and core BSE to obtain the corresponding dielectric tensors. These results build the basis for the calculation of the RIXS pathways, $t^{(1)}$, $t^{(2)}$, and $t^{(3)}(\omega_1)$, and the RIXS cross section $\frac{d^2\sigma}{d\Omega_2 d\omega_2}$ for a given choice of polarization directions of ingoing and outgoing light, $\mathbf{e}_1$ and $\mathbf{e}_2$, respectively. In the pumped case, a RT-TDDFT calculation is performed to obtain the time-dependent electronic occupations, which enter the BSE calculations and RIXS pathways as constraints.}
    \label{fig:workflow}
\end{figure}

\section{Application to graphite}\label{Sec:Results}

\subsection{Valence and core spectra}
\begin{figure}[h]
    \centering
    \includegraphics[width=0.45\textwidth]{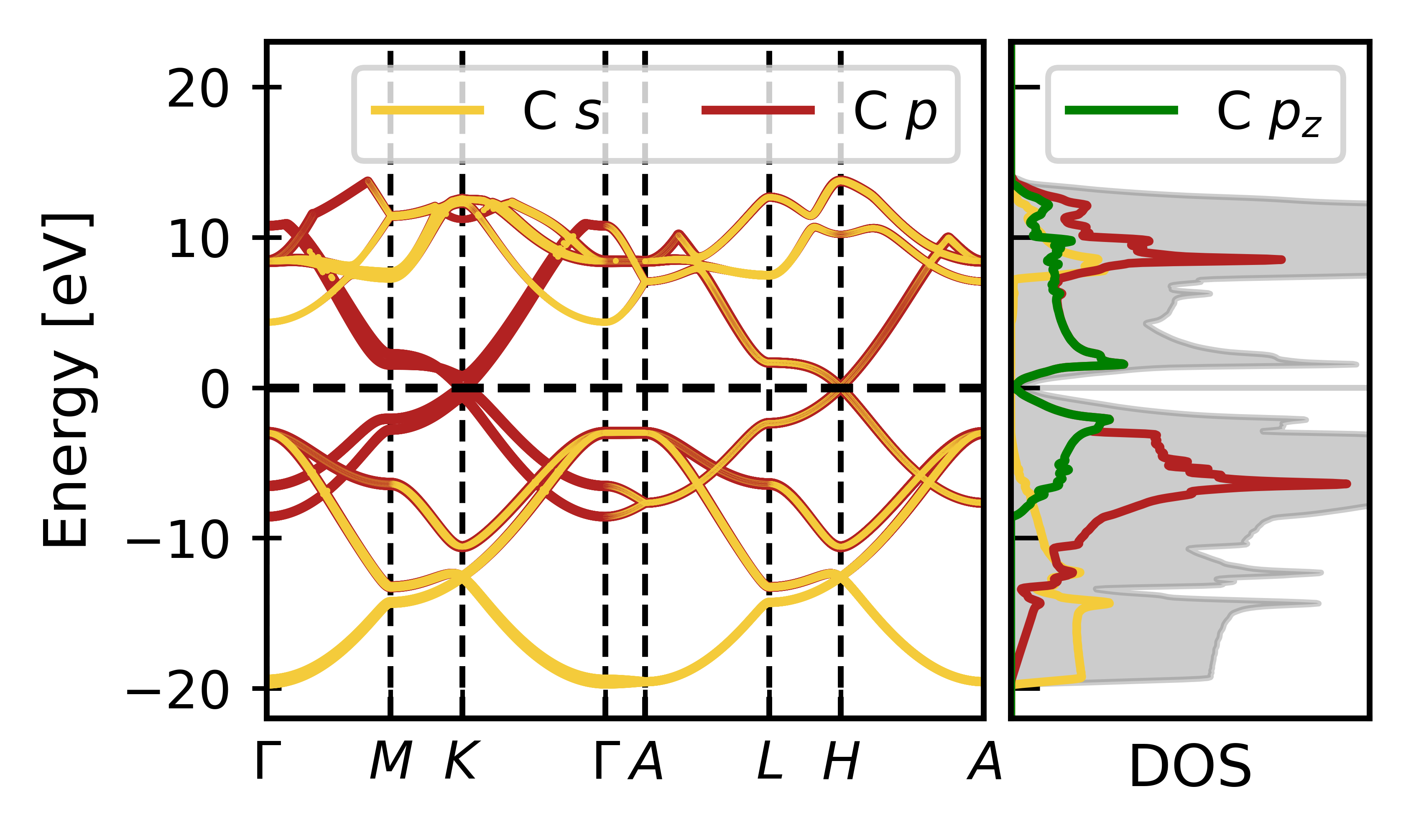}
    \caption{Band stucture (left) and density of states (right) of graphite, highlighting the band character. In the DOS, the pure $p_z$-contribution is shown in addition.}
    \label{fig:bs_dos}
\end{figure}

\begin{figure}[h]
    \centering
    \includegraphics[width= 0.45 \textwidth]{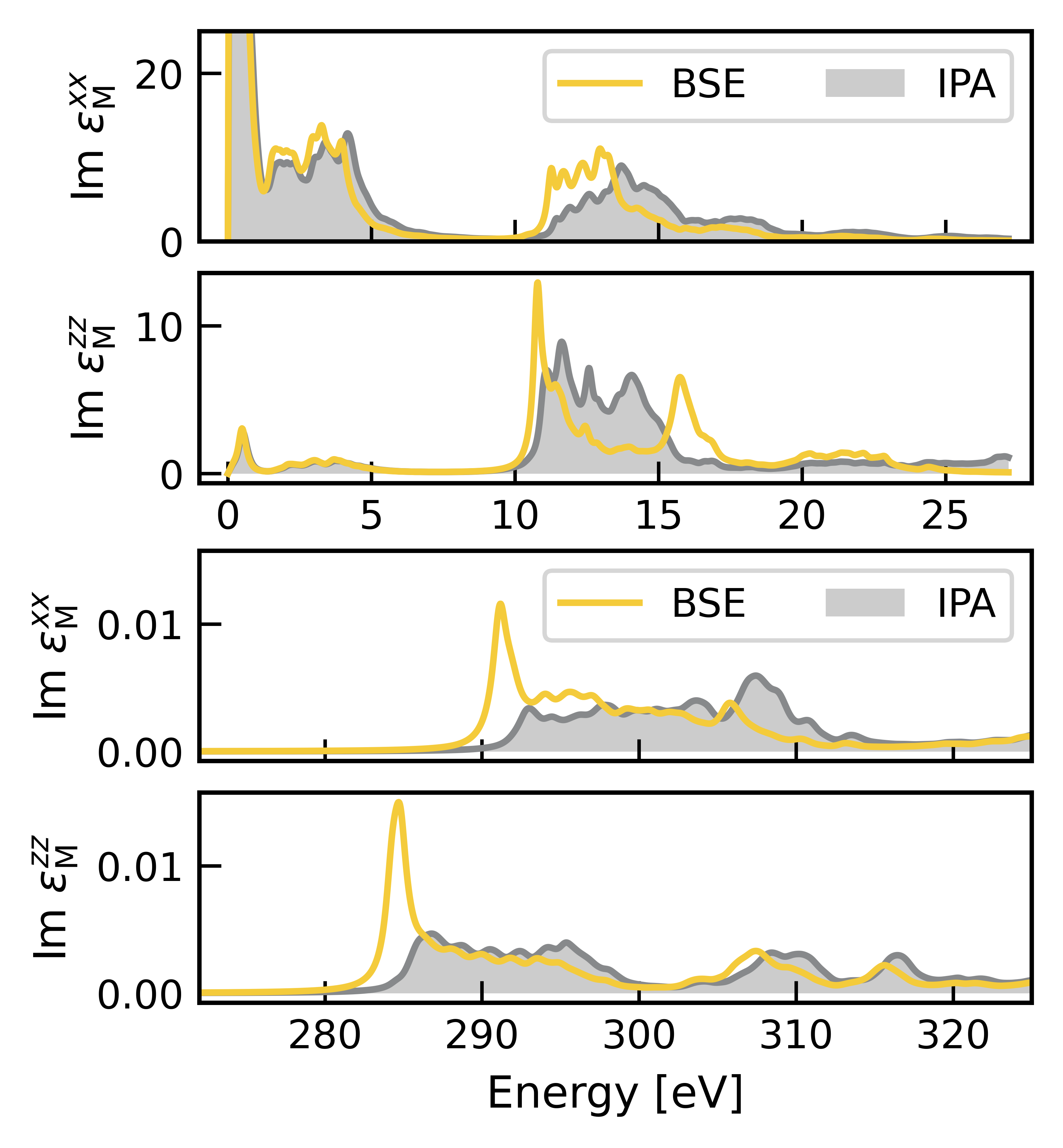}
    \caption{In-plane ($\epsilon_M^{xx}$) and out-of-plane ($\epsilon_M^{zz}$) components of the optical (top) and C K-edge core (bottom) spectra of graphite including (BSE, yellow) and excluding (IPA, gray) excitonic effects.}
    \label{fig:spec-stat}
\end{figure}

As an application of our theoretical framework, we study the carbon K-edge RIXS of graphite. Calculations of optical and core-level spectra of graphite have been discussed in the literature already. Since we will explore their interplay in terms of the RIXS cross section, we nevertheless first describe our results here, which are in good agreement with the previous results \cite{Trevisanutto2010,Marinopoulos2004}. The electronic band structure and density of states (DOS) are shown in Fig.~\ref{fig:bs_dos}, highlighting the $\pi$-like character of the states near the Fermi level, while $\sigma$-states are more strongly bound.
In \cref{fig:spec-stat}, we present the in-plane ($xx$) and the out-of-plane ($zz$) components of the imaginary part of the macroscopic dielectric tensor $\epsilon_M$, computed with the BSE. The IPA results are shown for comparison. 

The optical spectra already reflect the strong anisotropy of the electronic structure of graphite seen in \cref{fig:bs_dos} and discussed in the literature~\cite{Marinopoulos2004,Bassani1967}. The dominant low-energy structure in the range of approximately $0$--$5\,\si{eV}$ in the in-plane component, is associated with $\pi \to \pi^*$ transitions, particularly around the $K$-point. The fact that the main structures are already visible within the IPA indicates efficient metallic screening in these excitations. Excitonic effects become more pronounced only at higher energies between approximately $10$ and $15,\si{eV}$, where $\sigma \to \sigma^*$ contribute significantly. 

In contrast, the out-of-plane component exhibits considerably more pronounced exitonic effects. While there are only weak features below $10\,\si{eV}$, the dominant spectral structure between $10$--$15\,\si{eV}$  originates mainly from $\pi \to \sigma^*$ and $\sigma \to \pi^*$ transitions. The peak around $11\,\si{eV}$ is primarily associated with transitions between the highest occupied $\pi$ bands and the lowest unoccupied $\sigma^*$ bands. These transitions involve nearly parallel bands along the $M$-$K$-$\Gamma$ direction. At higher energies, around $14\,\si{eV}$, $\sigma \to \pi^*$ transitions become increasingly important, involving states mainly near the $M$ point of the Brillouin zone.

The characteristic peak structure of core-level absorption emerges only when electron–hole interactions are included. The pronounced peak around $284\,\si{eV}$ in the out-of-plane direction, originates from transitions into unoccupied $\pi^*$ states, while the peak around $292\,\si{eV}$, which is more prominent in the in-plane component, is dominated by transitions into $\sigma^*$ states. These findings are overall consistent with experimental results and literature data \cite{Shirley1998,Unzog2022}, although Olovsson and coworkers \cite{Olovson2019} report a noticeable influence of lattice vibrations, which are not accounted for in the present framework.

\subsection{Equilibium RIXS}
\begin{figure}[h]
    \centering
    \includegraphics[width=0.45\textwidth]{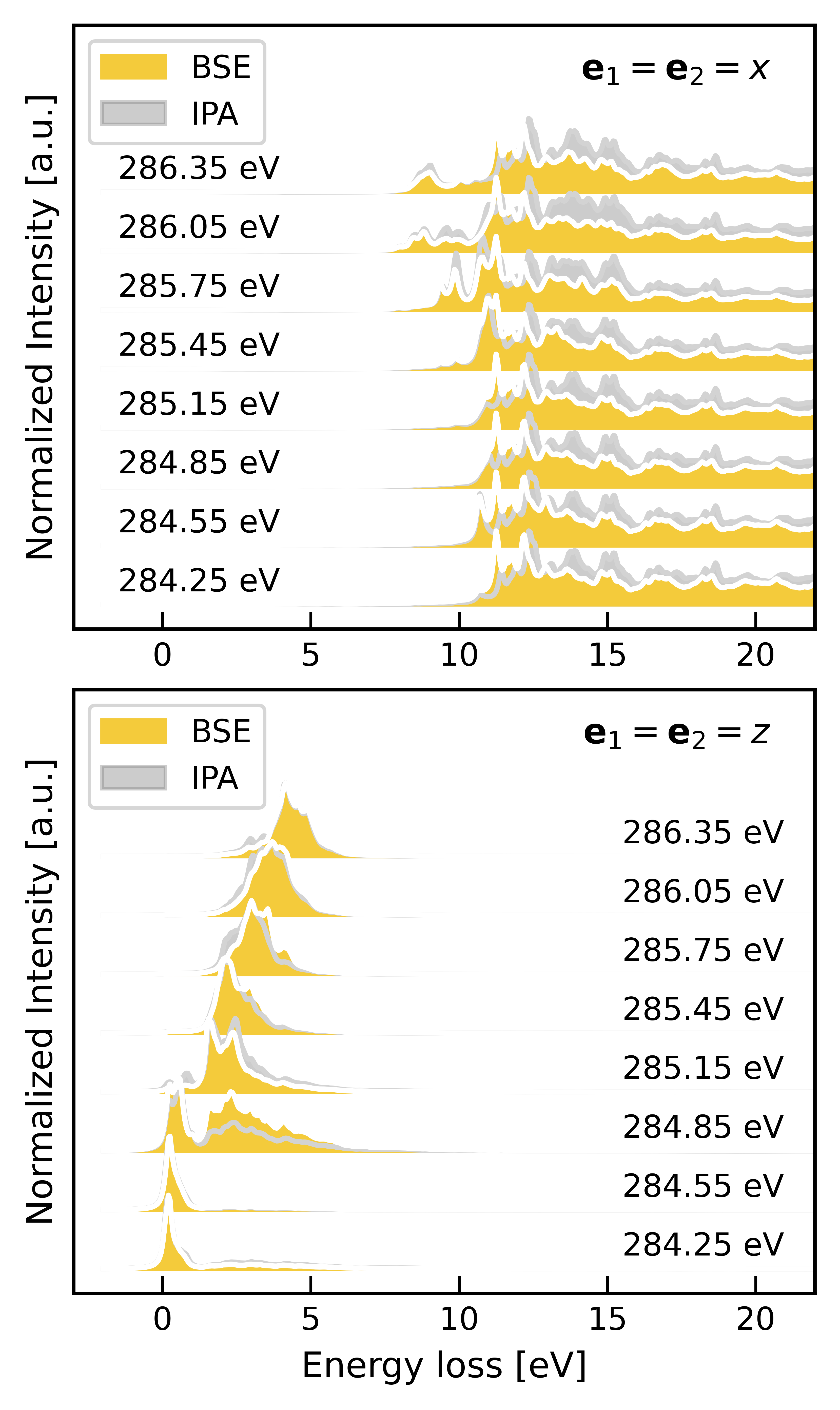}
    \caption{RIXS spectra for the carbon K-edge of graphite as a function of the energy loss from BSE (yellow) and the IPA (gray) for polarizations parallel to $x$ and $z$. Results for different excitation energies $\omega_1$ are offset in y-direction for clarity. }
    \label{fig:rixs-stat-xx-zz}
\end{figure}
\begin{figure*}[t]
    \centering
    \includegraphics[width=0.8\textwidth]{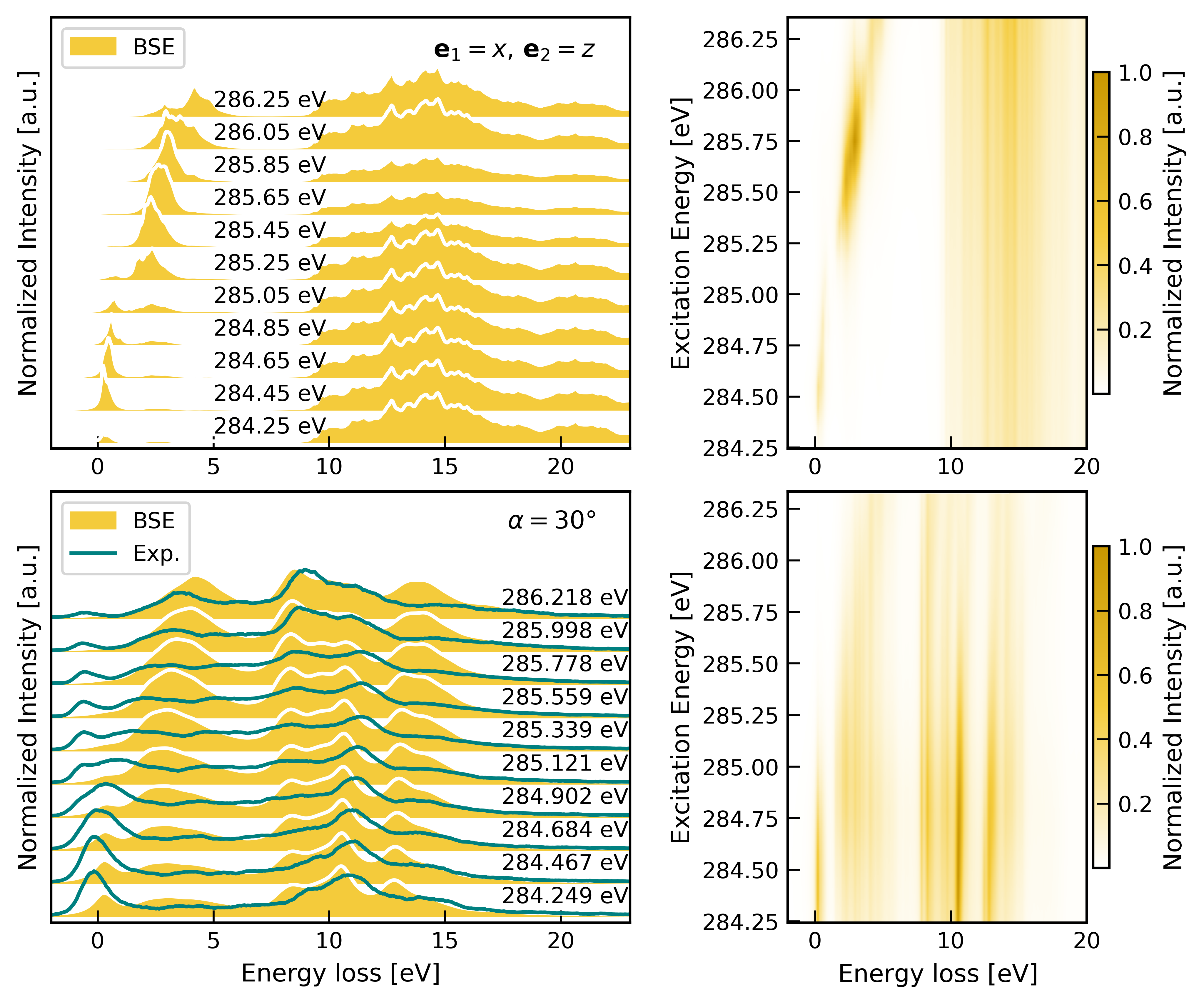}
    \caption{Left: RIXS spectra for the carbon K-edge of graphite as a function of the energy loss for incoming polarization parallel to $x$ and outgoing polarization parallel to $z$ (top) and for an incident angle of $30^\circ$ (bottom). Results for different excitation energies $\omega_1$ (yellow) are off-set in vertical direction for clarity. Experimental data are taken from Ref. \cite{Malvestuto2026} (teal). Right: Corresponding double-differential RIXS cross sections as a function of excitation energy $\omega_1$ and energy loss $\omega$ normalized to the maximum value.}
    \label{fig:rixs-stat-xz-30deg}
\end{figure*}

In \cref{fig:rixs-stat-xx-zz}, we display RIXS spectra as a function of the energy loss $\omega$ for different excitation energies $\omega_1$. Incoming and outgoing polarization directions are chosen parallel (upper panel) and perpendicular (lower panel) to the in-plane direction. Strikingly, the anisotropy, which was evident already in the core and optical spectra, becomes particularly strong in RIXS, where the intensities are pronounced either in one or the other region of energy loss, depending on the polarization direction. Comparison of results from BSE with the IPA affirms the non-negligible role of electron–hole interaction, despite the metallic nature of the material. While BSE results, including both the screened direct attraction and the counteracting local-field effects, show pronounced features already at lower $\omega_1$, the IPA spectra, neglecting both terms, start to exhibit significant intensity only at higher excitation energies $\omega_1$. 

To analyze the spectral features, we can relate them to those in  the optical and core-level excitation spectra, since the same anisotropic $\pi$- and $\sigma$-derived excitations also govern the RIXS response \cite{Carlisle1995}. For in-plane polarizations, the energy-loss spectrum is primarily characterized by $\sigma$-derived excitations at energies above $8\,\si{eV}$, reflecting transitions that involve deeper valence states of $\sigma$ character. In contrast, the out-of-plane polarizations are dominated by low energy-loss features involving $\pi$-type excitations. These contributions are associated with transitions near the $K$-point. Their spectral position shifts with incident photon energy, reflecting the dispersion of the underlying electronic structure. 

Having the full dielectric tensors for both core and valence excitations in hand, allows us to project out any arbitrary direction and thus investigate also mixed polarizations of incoming and outgoing light in RIXS. In \cref{fig:rixs-stat-xz-30deg}, this is exemplified with two cases. In the upper panel, we consider perpendicular polarizations of incoming and outgoing beam, which appears almost as a superposition of the spectral features discussed for the individual directions above. This could be expected, since these components exhibit significant intensities in different energy regions. 

In the lower panel, we show results for an incident angle of $30^\circ$, defined with respect to the out-of-plane direction, \ie the surface normal of the graphene layers. The scattering geometry is chosen such that the angle between incoming and outgoing beam is $90^\circ$, with  ($x$,$z$) being the scattering plane. This setup has been chosen, since it allows for direct comparison with experiment \cite{Malvestuto2026}. Since the polarization of the outgoing beam is not considered experimentally, we sample a set of uniformly distributed polarization vectors $\mathbf{e}_2$ and average over these spectra. The main  features are characterized by a non-linear mixture of $\pi$- and $\sigma$-derived contributions, obtained from a coherent mixing of both polarization components and thus revealing interference effects. In particular, angular-sensitive $\pi \to \pi^*$ transitions at low energy loss are enhanced. In contrast, for larger incident angles (not shown here), these transitions are suppressed. In the experimental data, the corresponding peak structure becomes distinguishable from the elastic peak around $0\, \si{eV}$ only at higher excitation energies $\omega_1$. At $285.12\,\si{eV}$ and above, a distinct peak emerges and shifts towards higher energy loss with increasing excitation energy. In contrast, the theoretical spectra exhibit the corresponding feature already at lower $\omega_1$, appearing at slightly higher energy loss, but showing similar dispersion. Note that full agreement between theory and experiment at very small energy losses is not expected, since the theoretical results do not include the elastic peak.

At higher energy loss, the spectra are dominated by $\sigma \to \sigma^*$ derived features, which exhibit only weak angular dependence due to the isotropy of the $\sigma$-orbitals. While the peak near $11\, \si{eV}$, corresponding to de-excitations from deeper valence $\, \sigma$-states, is dominant at low excitation energies, at higher excitation energies, the spectral weight shifts towards around $8\, \si{eV}$, which involves transitions from higher lying, less bound $\, \sigma$-states. At energy losses above $13\, \si{eV}$, a broader feature emerges, associated with deeper $\, \sigma$-states near the $M$-point. This feature exhibits a clear shift with increasing excitation energy, reflecting the dispersion of the underlying conduction bands. In the experimental data of Ref.~\onlinecite{Malvestuto2022}, its fine structure is not resolved and appears only as a broad shoulder extending up to $20\,\si{eV}$. A similar discrepancy has been reported for graphene~\cite{Zhang2012}. Earlier measurements of graphite, however, resolve a related feature at similar incident angle~\cite{Carlisle1995,Carlisle1999}. A possible explanation for the missing feature is self-absorption, \ie photon attenuation within the sample, which can suppress or distort spectral features, particularly in energy regions of strong absorption, depending on the experimental setup~\cite{Ament2011,Trevorah2019}. 

Overall, the main spectral features reproduce the peak positions observed in experiment. Remaining minor deviations may be attributed to electron–phonon coupling effects \cite{Dashwood2021}, which are not included in the present approach, and from performing BSE on top of DFT rather than using $GW$ quasi-particle energies.

\subsection{Pumped RIXS}
\begin{figure*}[t]
    \centering
    \includegraphics[width=0.8\textwidth]{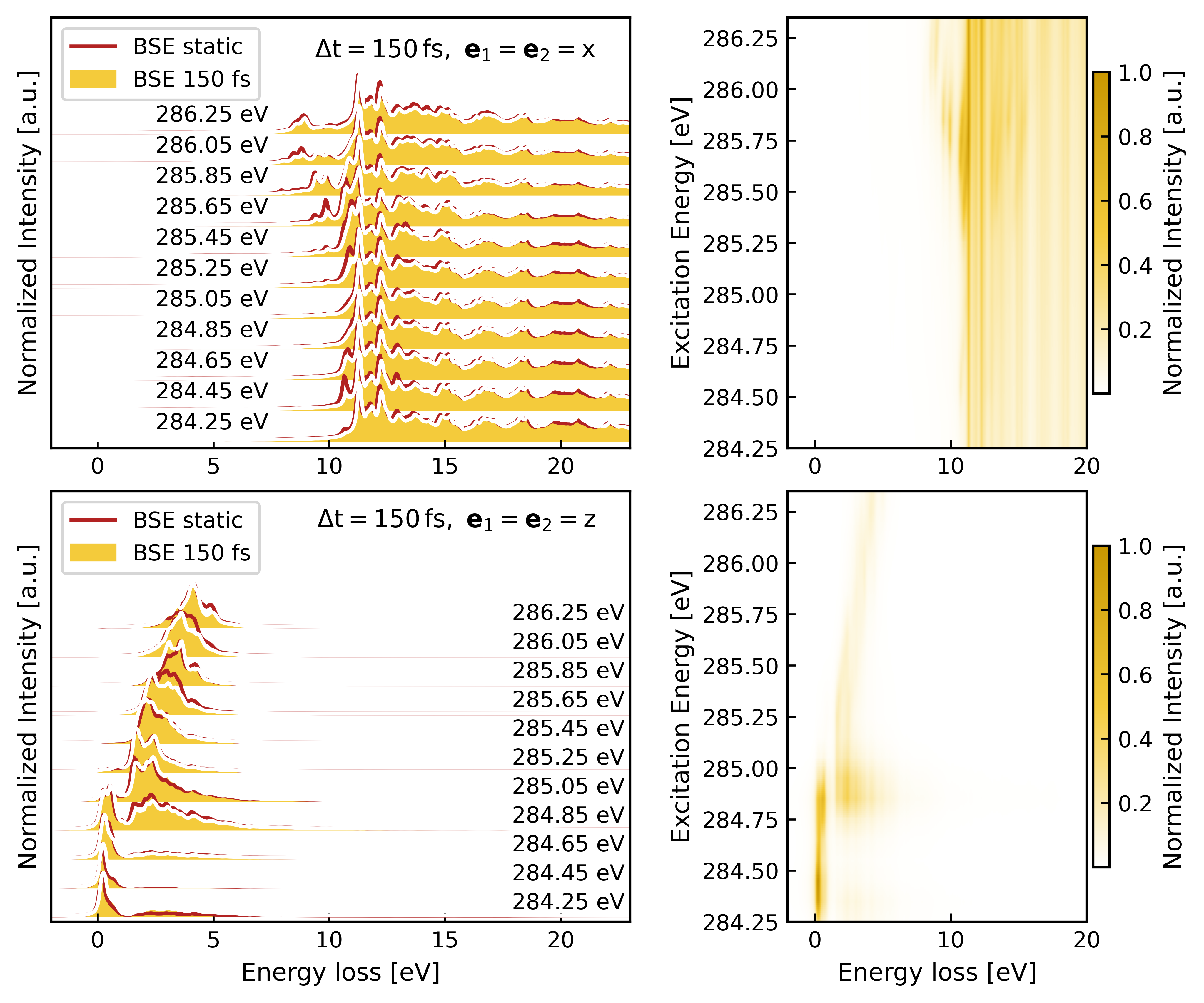}
    \caption{Left: Non-equilibrium RIXS spectra for the carbon K-edge of graphite as a function of the energy loss for polarizations parallel to $x$ and $z$, comparing static (red) and pumped (yellow) spectra. Results for different excitation energies $\omega_1$ are offset in vertical direction for clarity. The system is pumped for $45\,\si{fs}$ with a $266\,\si{nm}$ laser and a delay time of $150\,\si{fs}$ in the out-of-plane direction. Right: Corresponding double-differential RIXS cross sections for the pumped case as a function of excitation energy $\omega_1$ and energy loss $\omega$ normalized to the maximum value.}
    \label{fig:rixs-150fs-xx-zz}
\end{figure*}
\begin{figure*}[t]
    \centering
    \includegraphics[width=0.8\textwidth]{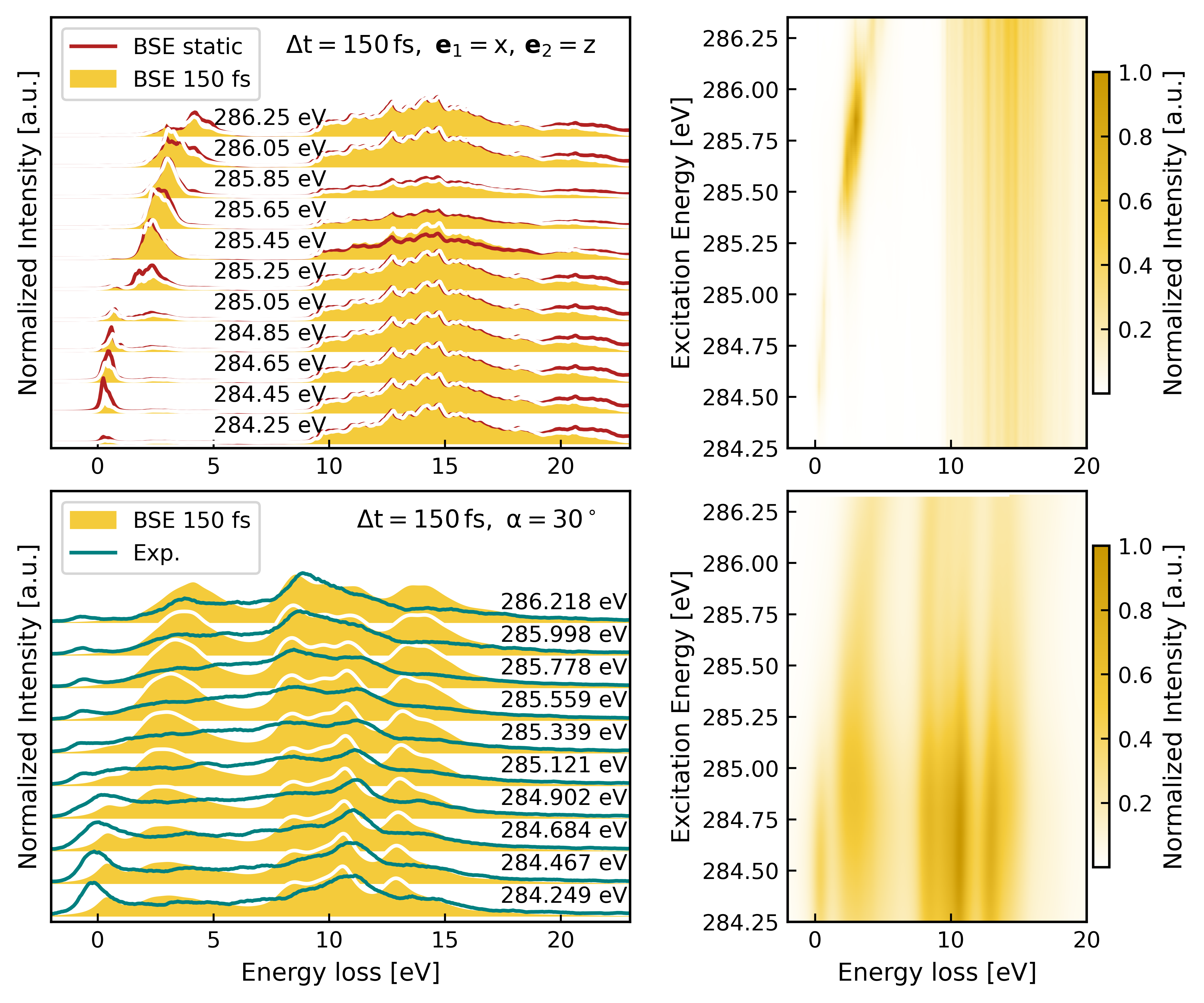}
    \caption{Left: Non-equilibrium RIXS spectra for the carbon K-edge of graphite, obtained by pumping for $45\,\si{fs}$ with a $266\,\si{nm}$ laser and a delay time of $150\,\si{fs}$ for different excitation energies $\omega_1$. Top: Pump pulse along $z$ in the out-of-plane direction with incoming polarization parallel to $x$ and outgoing polarization parallel to $z$, comparing static (red) and pumped (yellow) spectra. Bottom: Pump with an incident beam angle of $30^\circ$ and perpendicular polarization direction. Experimental data (teal) taken from \cite{Malvestuto2026}. Right: Corresponding double-differential RIXS cross sections for the pumped case as a function of excitation energy $\omega_1$ and energy loss $\omega$ normalized to the maximum value.}
    \label{fig:rixs-150fs-xz-30deg}
\end{figure*}

We now turn to the non-equilibrium case. In \cref{fig:rixs-150fs-xx-zz}, we display results for a fixed pump-probe delay time of $\Delta t=150\,\si{fs}$ and selected excitation energies. Incoming and outgoing polarizations are chosen in the same cartesian directions. Employing RT-TDDFT calculations, occupation numbers of the propagated system are obtained for a pump oriented in the out-of-plane direction to mimic the pump, with parameters chosen as listed in \cref{Sec:ComputationalDetails} to match the experimental setup in Ref. \cite{Malvestuto2026}. Overall, the spectra of the pumped system closely resemble those of the equilibrium configuration, indicating that the optical excitation induces only moderate modifications. Nevertheless, for polarization and pump along $z$, the spectra exhibit subtle modifications compared to the static low-energy-loss features, particularly at higher excitation energies. These changes consist mainly of small shifts and redistributions of relative spectral weights rather than the emergence of additional excitations and only occur in a region of relatively small absolute RIXS intensity.

For mixed polarization, we display tr-RIXS results for two setups in \cref{fig:rixs-150fs-xz-30deg}. In the upper panel, we again consider $z$ as the direction of the pump pulse, but the polarizations of incoming and outgoing photons are chosen perpendicular to each other. Overall, the results reproduce the findings observed for matching polarizations. Pump-induced effects are largely restricted to small changes in relative intensities. The most pronounced deviations occur in the low-energy-loss region, as in the case above. In the lower panel, we again adopt the geometry for an incident beam angle of $30^\circ$, with polarizations chosen as for the equilibrium case. In general, the non-equilibrium spectra closely resemble their equilibrium counterparts and deviations remain subtle. Nonetheless, the calculated spectra reproduce experimental features well up to energy losses of $12\,\si{eV}$, including a persistent shift in the $7$--$15\,\si{eV}$ region, which appears considerably less pronounced than at equilibrium. At higher energy losses, theory predicts an additional excitation feature at high excitation energies, already visible at the equilibrium level, which corresponds to transitions from deep $\sigma$-states. 

Overall, time-resolved spectral changes are well captured. While they are more pronounced for mixed cartesian directions, for an incident angle they of $30^\circ$ they are more subtile. Agreement with experiment as present in the equilibrium case remains good with the main spectral features being reproduced. Distinctions between theory and experiment remain constant across the considered excitation energies and have been attributed to electron-phonon coupling \cite{Malvestuto2026}. 

\section{Computational details}
\label{Sec:ComputationalDetails}

All calculations were carried out with the all-electron full-potential package \exciting \cite{Exciting2014,BSEexciting,Exciting2026}. Ground-state properties, including structural relaxation, were obtained from DFT using the GGA-PBE exchange–correlation functional with Tkatchenko–Scheffler van der Waals corrections~\cite{TSvdW}. For the ground-state calculations, a 24$\times$24$\times$9 $\mathbf{k}$grid was used for Brillouin-zone integrations, and a basis-set size with an $R^{\mathrm{max}}_{\mathrm{MT}} |G + k|_{\mathrm{max}}$ value of 8 was chosen, corresponding to a planewave cutoff of 6.66 (muffin-tin radius of $R_{\mathrm{MT}}=1.2 \, a_0$) for the Kohn-Sham orbitals. The BSE calculations were performed on a 12$\times$12$\times$4 $\mathbf{k}$ grid with a planewave cutoff of $|G + q|_{\mathrm{max}} =3.5\, a_0^{-1}$. All 8 occupied valence bands and 15 conduction bands were considered in the optical spectra. For the X-ray spectra, transitions into 40 conduction bands were included, and a scissors shift of $18.5 \, \si{eV}$ was applied. A Gaussian broadening of $0.15\, \si{eV}$ ($0.49\, \si{eV}$) was applied for optical (core) spectra. 

For the static RIXS calculations, the 64,000 lowest-energy valence and 20,000 carbon K-edge excitations were considered. For comparison with experiment, a lifetime broadening of $0.5\, \si{eV}$ was estimated as a convolution of core lifetime and instrumental resolution in the experiment \cite{Chernenko2021}. For all other calculations, a lifetime broadening of $0.1\, \si{eV}$ was used, matching the experimental core-hole lifetime \cite{Harada2004,Nicolas2012}. 

For the pumped case, RT-TDDFT calculations were carried out to simulate the system excited by a laser pulse with a wavelength of $266 \, \si{nm}$ for a duration of $45 \, \si{fs}$ creating a pump fluence of $30 \, \si{mJ/cm^2}$ in the sample. Non-equilibrium electronic occupation numbers were calculated for delay times of up to $300\,\si{fs}$ in time-steps of $0.25\,\si{au}$ as input for non-equilibrium BSE calculations~\cite{Rossi2025,qiao2026},  applying it to both the optical and X-ray absorption spectra. Parameters were chosen to match experimental data \cite{Malvestuto2022,Malvestuto2025}. For polarization along cartesian axes, the pump was considered in $z$-direction. For comparison with experiment at incident beam angle of $30^\circ$, relative to the normal of the graphene layers in the ($x$,$z$) plane. For non-equilibrium RIXS calculations, all valence and core-level excitations in the energy ranges of $0-25\,\si{eV}$ and $280-300\,\si{eV}$, respectively, were taken into account. 

Input and output files of all calculations are available on NOMAD \cite{Scheidgen2023}, DOI https://doi.org/10.17172/nomad.cd5k-mb74.

\section{Conclusions}

We have presented a compact approach to obtain polarization- and time-resolved RIXS spectra. Being based on the Bethe-Salpeter equation of many-body perturbation theory, our method provides a general and fully parameter-free pathway to compute RIXS spectra in complex materials, from metals to large-gap semiconductors for arbitrary scattering geometries. In combination with RT-TDDFT, it also accounts for pumped, \ie time-dependent scenarios. Using the carbon K-edge of the semimetal graphite as a representative example, we have demonstrated this approach for static and non-equilibrium RIXS spectra. Our results reflect the characteristic anisotropy of graphite, showing up in distinct $\pi$- and $\sigma$-derived features and their evolution with incident photon energy, in excellent agreement with experiment. 

\section*{Acknowledgment}
We thank Lu Qiao and Ronaldo Rodrigues Pela for providing their implementation of the non-equilibrium Bethe–Salpeter equation formalism in \exciting prior to publication.

\bibliography{bib}
\end{document}